\documentclass[11pt]{article}

\usepackage[english]{babel}
\usepackage[letterpaper,top=1in,bottom=1in,left=1in,right=1in,marginparwidth=0.75in]{geometry}

\usepackage[version=4]{mhchem} 
\usepackage{siunitx}
\usepackage{graphicx}
\usepackage{dcolumn}
\usepackage{bm}
\usepackage{setspace}
\usepackage{amsmath}
\usepackage{fixmath}
\usepackage{graphicx}
\usepackage{amssymb}
\usepackage{mathtools}
\usepackage{cite}
\usepackage{algorithmic}
\usepackage{algorithm}
\usepackage{microtype}
\usepackage{graphicx}
\usepackage{subfigure}
\usepackage{booktabs} 

\usepackage{amsmath}
\usepackage{amssymb}
\usepackage{mathtools}
\usepackage{amsthm}
\usepackage{color}

\DeclareMathOperator*{\argmin}{arg\,min}

\title{Distributed Tomographic Reconstruction with Quantization}
\author{Runxuan Miao$^{\dagger}$, Selin Aslan$^{\ddagger}$, Erdem Koyuncu$^{\dagger}$,\, Do\u ga G\" ursoy\footnote{E-mail: dgursoy@anl.gov} \\ 
$^{\dagger}$Department of Electrical and Computer Engineering, University of Illinois Chicago\\ $^{\ddagger}$Department of Mathematics, Ko\c c University \\ 
*Advanced Photon Source, Argonne National Laboratory}

\begin{document}
\doublespacing
\maketitle

\begin{abstract}
Conventional tomographic reconstruction typically depends on centralized servers for both data storage and computation, leading to concerns about memory limitations and data privacy. Distributed reconstruction algorithms mitigate these issues by partitioning data across multiple nodes, reducing server load and enhancing privacy. However, these algorithms often encounter challenges related to memory constraints and communication overhead between nodes. In this paper, we introduce a decentralized Alternating Directions Method of Multipliers (ADMM) with configurable quantization. By distributing local objectives across nodes, our approach is highly scalable and can efficiently reconstruct images while adapting to available resources. To overcome communication bottlenecks, we propose two quantization techniques based on K-means clustering and JPEG compression. Numerical experiments with benchmark images illustrate the tradeoffs between communication efficiency, memory use, and reconstruction accuracy.
\end{abstract}

\section{Introduction}

Tomographic reconstruction is the process of determining an object's internal structure from a series of projection imagess, known as sinograms, recorded by passing beams such as x-rays or electrons through the object as it rotates \cite{kak2001principles}. This non-invasive and rapid imaging technique is widely used in medical imaging, where Computed Tomography (CT) generates detailed cross-sectional images of the body, while Positron Emission Tomography and Single-Photon Emission Computed Tomography use radioactive tracers to map metabolic activity and blood flow \cite{wernick2004emission}. In materials science, methods such as Micro-CT and Nano-CT provide high-resolution imaging of material structures, allowing for analysis of complex materials \cite{withers2021x}.

\begin{figure}
\centering
  \includegraphics{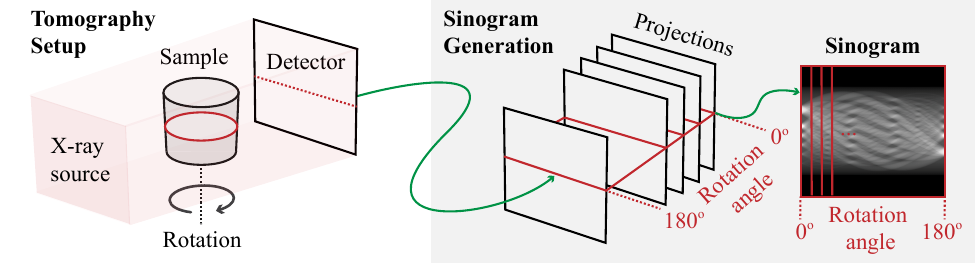}
  \caption{Illustration of the tomographic data acquisition setup. The sample is illuminated with x-rays and rotated while projections are captured at various angles, typically ranging from \SI{0}{\degree} to \SI{180}{\degree}. These projections, recorded by the detector, are compiled into a sinogram that are correlated with the intensity of x-rays transmitted through the object at each angle.}
  \label{fig1}
\end{figure}

Fig.~\ref{fig1} illustrates the fundamental principle of tomography. The technique involves illuminating an object with x-rays, producing its silhouette on a pixel array detector positioned downstream. As the x-rays pass through the object, their intensity decreases due to absorption and scattering caused by interactions with the material. Each pixel records this reduction in intensity along the beam path, capturing the cumulative attenuation properties of the sample. By collecting these measurements from multiple pixels and at various rotation angles, an image called a sinogram is generated. Tomographic reconstruction algorithms then use the sinogram data to reconstruct the internal distribution of attenuation-related properties, such as density or chemical composition, resulting in a detailed representation of the object’s internal structure.

The volume and rate of tomography data are rapidly increasing, particularly in synchrotron-based imaging \cite{de2018tomobank}. Synchrotrons provide exceptionally high brightness, often many orders of magnitude greater than that of conventional medical CT systems. This increased brightness facilitates faster imaging and higher-resolution scans, enabling significantly shorter acquisition times. For example, current detectors typically achieve resolutions of up to 4 gigapixels, and can generate several terabytes of data within seconds \cite{garcia2021tomoscopy}. With ongoing upgrades to synchrotron sources, further improvements in these metrics are anticipated in the near future. This surge in data generation underscores the need for efficient reconstruction algorithms that can manage large-scale datasets while effectively addressing imaging artifacts and noise.

Current algorithms for large-scale tomography data predominantly rely on analytic methods due to their speed and computational efficiency. Techniques such as Filtered Backprojection \cite{kak2001principles} and its variants including Feldkamp-Davis-Kress \cite{feldkamp1984practical} and Katsevich’s algorithm \cite{katsevich2002theoretically} are widely utilized because they can produce rapid reconstructions with relatively low computational demands. These methods leverage mathematical transformations, particularly fast Fourier transforms, to convert sinogram data into images efficiently. The introduction of Graphics Processing Units has further improved the performance of these transformations \cite{andersson2016fast}. Despite their advantages, analytic methods often encounter challenges when dealing with noisy or incomplete data and can be less effective in non-standard acquisition geometries \cite{koyuncu2022centroidal}. 

In contrast, iterative methods offer a more flexible and robust approach to tomographic reconstruction. Early iterative techniques, such as the Algebraic Reconstruction Technique \cite{gordon1970art}, Simultaneous Algebraic Reconstruction Technique \cite{andersen1984simultaneous}, and Simultaneous Iterative Reconstruction Technique \cite{gilbert1972iterative}, excel in handling underdetermined systems and are relatively straightforward to implement. More modern approaches focus on setting and minimizing cost functions through iterative refinements, employing various optimization tools, including Krylov subspace methods like Conjugate Gradient Least Squares \cite{hestenes1952methods}, gradient descent \cite{nesterov2013introductory}, and Newton's method \cite{nocedal2006numerical}. Additionally, statistical techniques such as Maximum Likelihood Expectation Maximization \cite{lange1984reconstruction} are particularly effective in low-dose applications, relying on realistic noise models to enhance reconstruction quality. Despite these strengths, applying iterative methods to large-scale data remains challenging due to their slower convergence rates and difficulties with parallelization, which can limit their practical use in processing the large-scale datasets. Consequently, as dataset sizes continue to grow, the increased demands on memory and processing power make storage and reconstruction on central servers increasingly impractical \cite{10.1145/3653981, 10683880}.

To address the increasing computational challenges in tomographic reconstruction, parallel computing has emerged as a leading solution that leverages advancements in modern hardware. Multi-core processors, graphical processing units (GPUs), and parallel computing frameworks have significantly accelerated many tasks within this field \cite{yu2006gpu, wang2007performance}. Distributed-memory implementations enable faster reconstructions by distributing data and computations across multiple processing units. Many of these implementations utilize slice-based parallelization, where the data volume is segmented into sinograms that can be independently reconstructed \cite{bicer2015rapid}. In contrast, in-slice parallelization, which involves parallelization within a single slice or sinogram, has been less extensively explored, with only a few notable exceptions \cite{cui2013distributed, chen2021scalable, xu2020enhanced, hidayetouglu2020petascale, hidayetouglu2021memxct}. While these methods are scalable, they often exhibit a high degree of specialization and lack flexibility in adapting solver types based on the specific problem at hand. Moreover, their generalizability is limited, necessitating configuration adjustments whenever the computing environment changes. This presents particular challenges in rapidly evolving computational landscapes, where new systems may offer varying communication, memory, and processing capabilities \cite{9478829, 10138654}.

Recent advancements in distributed computing and optimization algorithms offer significant opportunities to tackle the complexities inherent in large-scale problem solving for tomographic reconstruction \cite{buurlage2019geometric, biguri2020arbitrarily}. The Alternating Directions Method of Multipliers (ADMM) \cite{boyd2011distributed} has emerged as a particularly effective framework in this context. By decomposing complex problems into smaller, manageable sub-problems that can be addressed locally at each processing node, ADMM simplifies the overall computational process. Its flexibility allows for the integration of various imaging tasks, such as phase retrieval \cite{aslan2019joint, nikitin2019photon}, denoising \cite{nikitin2021distributed}, deblurring \cite{majee2022codex}, and regularization \cite{aslan2021joint, barutcu2021limited}, within the tomographic reconstruction framework. Despite the ongoing development of its theoretical foundations \cite{sun2021scalable}, ADMM's compatibility with existing systems enhances its practical implementation. However, challenges persist. Effective load balancing in terms of memory, communication, and processing across distributed nodes is difficult to achieve, which can lead to inefficiencies. Additionally, although there is a substantial body of literature on ADMM, issues such as slow convergence rates and scalability challenges arise, particularly due to the need for consensus among solutions or the requirement for an all-gather operation across multiple nodes \cite{sridhar2020distributed}. As the size of the problems grows, addressing these bottlenecks, especially concerning high data bandwidth requirements among nodes, will be critical for maximizing ADMM’s capabilities in real-world applications.

To address these challenges, we propose an ADMM framework with configurable quantization to improve communication efficiency across nodes. By compressing the data shared among local nodes, our approach reduces the communication load, ensuring scalability and alleviating bottlenecks for nodes dependent on data from others. We focus on K-means and JPEG methods, which are computationally efficient, to enhance data transmission. Additionally, to lessen the burden on a centralized server, which is a limitation of traditional ADMM, we adopt a decentralized ADMM (dADMM) approach that avoids the all-gather operation by implementing a partial communication pattern among nodes. Our model outlines the algorithmic requirements for solving the reconstruction problem across multiple nodes with specific computational resources. Numerical experiments explore the trade-off between reconstruction accuracy and runtime performance under various measurement noise conditions. The results demonstrate the potential of decomposing tomographic problems into low-bandwidth nodes, effectively reducing memory demands and computational complexity on the central server while offering insights into the balance between reconstruction accuracy and runtime performance.

\section{Methods}

In this section, after providing the preliminaries in tomographic imaging, we begin by presenting the baseline approach, referred to as Centralized Tomographic Reconstruction (CTR). Next, we extend this to a distributed setting with ADMM. Finally, we introduce the quantization technique to improve the network communication efficiency.

\subsection{Background in Tomographic Reconstruction}

The process of tomographic reconstruction begins with the acquisition of sinogram (projection data), where each sinogram represents the integral of the object’s attenuation coefficients along the path of the x-ray beam. Mathematically, this can be described by the x-ray transform, which converts a 2D function (the cross-sectional image of the object) into a set of 1D projections. 
\begin{equation}
    \mathcal{P}(\theta, t) = \int_{-\infty}^{\infty} u\left(t \cos(\theta) - s \sin(\theta), \, t \sin(\theta) + s \cos(\theta)\right) \, ds
\end{equation}
where $\mathcal{P}$ is the x-ray transform of the 2D attenuation function $u$, $\theta$ is the angle of the x-ray beam relative to a fixed axis, $t$ is the perpendicular distance from the origin to the line along which the x-ray is taken, and $s$ is the variable of integration along the line parameterized by 
$t$ and $\theta$.

To reconstruct an object's internal structure from its x-ray projections, we discretize the continuous x-ray transform into a linear system of equations. This process involves dividing the 2D object into a grid of pixels and considering the x-ray paths through each pixel. The system matrix $P$ is formed by discretizing the x-ray transform integral:
\begin{equation}
d_j = \sum_{i=1}^n P_{ji} u_i
\end{equation}
where, $d_j$ is the $j$-th projection data (corresponding to a particular x-ray path), $u_i$ is the attenuation coefficient of the $i$-th pixel, $P_{ji}$ is the length of the intersection between the $j$-th x-ray path and the $i$-th pixel. By stacking all the projection equations together, we obtain the complete linear system:
\begin{equation}
\begin{bmatrix}
d_1 \\
d_2 \\
\vdots \\
d_l
\end{bmatrix}
=
\begin{bmatrix}
P_{11} & P_{12} & \cdots & P_{1n} \\
P_{21} & P_{22} & \cdots & P_{2n} \\
\vdots & \vdots & \ddots & \vdots \\
P_{l1} & P_{l2} & \cdots & P_{ln}
\end{bmatrix}
\begin{bmatrix}
u_1 \\
u_2 \\
\vdots \\
u_n
\end{bmatrix}
\end{equation}
or more compactly:
\begin{equation}
d = P u
\label{forward}
\end{equation}
where, $l$ is the total number of projections (x-ray paths), and $n$ is the total number of pixels in the image. Solving this system of equations enables the reconstruction of the vector $u$, which characterizes the internal structure of the object and will be further detailed in the following section.

\subsection{Centralized Tomographic Reconstruction}

To solve for the $u$ in Eq.~\ref{forward}, we formulate the following CTR problem:
\begin{align}
\label{eq:single_rec}
    & \min_{u} \frac{1}{2} \| P u - d \|_2^2,
\end{align}
where $d \in \mathbb{R}^{l}$ represents the sinogram of an image, $P \in \mathbb{R}^{l \times n}$ denotes the x-ray transform, and $u \in \mathbb{R}^n$ is the unknown image to be reconstructed. In this paper, we employ a standard gradient descent method, outlined in Algorithm~\ref{alg:gd}, as our baseline approach for solving the CTR problem.

\begin{algorithm}[b]
    \caption{Gradient Descent to solve the CTR problem}
	\label{alg:gd}
	\begin{algorithmic}[1]
		\renewcommand{\algorithmicrequire}{\textbf{Input:}}
		\renewcommand{\algorithmicensure}{\textbf{Output:}}
		\REQUIRE $P \in \mathbb{R}^{l \times n}$ and $d \in \mathbb{R}^{l}$, learning rate (step size) $\eta$, and the number of iterations $E$.
        \ENSURE  $u \in \mathbb{R}^n$.
        \STATE{Initialize $u \in \mathbb{R} ^n$ with all zeros.}
        \FOR{$i = 1, \ldots, E$}
        \STATE{$\Delta_{u} = P^T(P u - d)$}
        \STATE{$u \leftarrow u - \eta \Delta_{u}$}
        \ENDFOR
	\end{algorithmic}
\end{algorithm}

\subsection{Distributed Tomographic Reconstruction with ADMM}

We extend the centralized scheme to a distributed setting using ADMM. Consider a distributed system with $M$ nodes, where each local node $m$ stores a portion of the data, denoted as $d_m \in \mathbb{R}^{l^1}$, with $l^1 = l/M$. While the amount of data per node can be adjusted for heterogeneous nodes, for simplicity, we assume in this paper that all nodes are identical, splitting the data evenly. We partition the tomographic data across multiple nodes based on different sets of rotation angles. The projection matrix for each node is represented as $P_m \in \mathbb{R}^{l^1 \times n}$. The reconstruction problem is then formulated as a constrained optimization problem:
\begin{align}
\label{eq:dis_prob}
    & \min_{x, u_1, \ldots , u_m} \frac{1}{2} \sum_{m=1}^M \| P_m u_m - d_m \|_2^2 \\
    & \quad \text{subject to} \quad u_m = x, \quad m=1,2,\ldots,M \nonumber
\end{align}
where, $u_1, \ldots, u_M \in \mathbb{R}^n$ are introduced as auxiliary variables to facilitate the relaxation of the problem. In simpler terms, we introduce additional unknown images, $u_1, \ldots, u_M \in \mathbb{R}^n$, which will be solved alongside $x$. As we will show next, this approach enables the distribution of solutions across multiple nodes, allowing each node to work on its respective solution while ensuring that all images converge to be identical. Consequently, a certain level of communication between nodes is necessary to share progress and ensure alignment of solutions, allowing for effective collaborative problem-solving.

To solve Eq.~\ref{eq:dis_prob}, we utilize the following augmented Lagrangian which forms as the basis of the ADMM approach:
\begin{align}\label{Eq:AugLag}
\mathcal{L}_\rho(u_1, \ldots, u_M, x, \lambda_1,  \ldots, \lambda_M) = \sum_{m=1}^M\bigg(\frac{1}{2} \Vert P_m u_m - d_{m} \Vert_2^2  +  \lambda_m^T(u_m - x) 
 + \frac{\rho}{2}\Vert u_m - x \Vert_2^2\bigg),
\end{align}
where, $\lambda_1, \cdots, \lambda_M \in \mathbb{R}^n$ are dual variables, and $\rho$ represents the penalty parameter. In an ADMM framework, the augmented Lagrangian, or the cost function, involves multiple unknowns that can be assigned to available computational resources and solved alternately. This iterative process updates each variable sequentially, leading to the final solution using the ADMM algorithm \cite{boyd2011distributed} as follows:
\begin{align}
\label{eq-compact:solve_u}
u_m^{k+1} &= \argmin_{u_m} \mathcal{L}_\rho\left(u_m, x^{k}, \lambda_m^{k}\right),
& m=1, 2, \ldots, M, \\ 
\label{eq-compact:solve_x}
x^{k+1} &= \argmin_x \mathcal{L}_\rho\left(u_1^{k+1},  \ldots, u_m^{k+1}, x, \lambda_1^{k},  \ldots, \lambda_m^{k}\right), \\ 
\label{eq-compact:solve_lambda}
\lambda_m^{k+1} &= \lambda_m^k + \rho \left(u_m^{k+1}- x^{k+1}\right),
& m=1, 2, \ldots, M,  
\end{align}
where $k$ denotes the iteration number in ADMM. In this algorithm, Eq.~\ref{eq-compact:solve_u} involves $M$ distinct sub-problems, which can be efficiently distributed across $M$ nodes for parallel processing. The updated variables are then aggregated at a central node to solve the problem in Eq.~\ref{eq-compact:solve_x}. Subsequently, the dual variables are updated using all $M$ nodes by using Eq.~\ref{eq-compact:solve_lambda}. However, the updated $x$ must be transmitted back to each of these nodes to complete one ADMM iteration.



\subsection{Decentralized ADMM}

The standard ADMM, when applied to solve Eqs.~\ref{eq-compact:solve_u}, \ref{eq-compact:solve_x}, and \ref{eq-compact:solve_lambda} cyclically, encounters scalability challenges. This arises from the need to gather $u_m$ and $\lambda_m$ values from all nodes during the solution of Eq.~\ref{eq-compact:solve_x}, resulting in linearly increasing communication overhead as the number of nodes grows. To address this issue, we partition the variable $x$ into $M$ segments, with each node responsible for updating its specific segment, denoted as $x[m]$. The complete image $x$ is reconstructed by concatenating these segments, i.e., $x = \text{conc}(x[1], \cdots, x[M])$. This approach relaxes the strict updating of $x$ in each node, leading to the reformulation of the ADMM problem as follows:
\begin{align}
\label{eq-compact:solve_u2}
u_m^{k+1} &= \argmin_{u_m} \mathcal{L}_\rho\left(u_m, x^{k}, \lambda_m^{k}\right),
& m=1, 2, \ldots, M, \\ 
\label{eq-compact:solve_x2}
x[m]^{k+1} &= \argmin_{x[m]} \mathcal{L}_\rho\left(u_m^{k+1}, x[m], \lambda_m^{k}\right), & m=1, 2, \ldots, M, \\ 
\label{eq-compact:conc2}
x^{k+1} &= \text{conc}(x[1]^{k+1}, \cdots, x[M]^{k+1}), \\
\label{eq-compact:solve_lambda2}
\lambda_m^{k+1} &= \lambda_m^k + \rho \left(u_m^{k+1}- x^{k+1}\right),
& m=1, 2, \ldots, M,  
\end{align}
In this decentralized ADMM (dADMM) approach, after each node computes its assigned segment, $x$ can be reconstructed globally by exchanging these segments among the nodes. Importantly, this method ensures that communication overhead remains constant as the number of nodes increases. Because as more nodes are added, each segment's size decreases proportionally, leading to consistent communication demands and ensuring the scalability of the algorithm. 

\subsection{Implementation}
\label{sec:dis_admm}

By explicitly deriving the update rules for dADMM, we substitute the augmented Lagrangian from Eq.~\ref{Eq:AugLag} into the sub-problems defined in Eqs.~\ref{eq-compact:solve_u2}-  \ref{eq-compact:solve_lambda2} as follows
\begin{align}
\label{eq:solve_u}
u_m^{k+1} &= \argmin_{u_m} \bigg(\frac{1}{2} \Vert P_m u_m - d_{m} \Vert_2^2  + \frac{\rho}{2}\Vert u_m - x^k +  \lambda_m^{k}/ \rho \Vert_2^2\bigg),
& m=1, 2, \ldots, M, \\ 
\label{eq:solve_x}
x[m]^{k+1} &=\argmin_{x[m]} \bigg(\frac{\rho}{2}\Vert u_m^{k+1} - x[m] +  \lambda_m^{k}/ \rho \Vert_2^2\bigg), &  m=1, 2, \ldots, M,  \\
\label{eq:conc}
x^{k+1} &= \text{conc}(x[1]^{k+1}, \cdots, x[M]^{k+1}), \\
\label{eq:solve_lamda}
\lambda_m^{k+1} &= \lambda_m^k + \rho \left(u_m^{k+1}-x^{k+1}\right),
& m=1, 2, \ldots, M. 
\end{align}
The dADMM iterations start by initializing all variables and assigning respective parts of data to each node. $M$ problems in Eq.~\ref{eq:solve_u} take the form of standard tomographic reconstruction problems with Tikhonov regularization \cite{tikhonov1977illposed}, where $\rho$ serves as the regularization parameter balancing data fidelity and regularization. The updates for $u_1, \ldots, u_M$ can be independently computed on each local node using gradient descent, detailed in Algorithm~\ref{alg:dis_gd_to_solve_u}. Note that it is not necessary to solve the sub-problems exactly to ensure the convergence of the dADMM algorithm ~\cite[Th.8]{eckstein1992douglas}. Hence, we approximately solve the problems by using several gradient descent iterations for rapid convergence as also suggested in \cite{boyd2011distributed}.

\begin{algorithm}[b]
	\caption{Gradient Descent to solve Eq.~\ref{eq:solve_u}}
	\label{alg:dis_gd_to_solve_u}
	\begin{algorithmic}[1]
	\renewcommand{\algorithmicrequire}{\textbf{Input:}}
		\renewcommand{\algorithmicensure}{\textbf{Output:}}
		\REQUIRE $P_m \in \mathbb{R}^{l^1 \times n}$ and $d_m \in \mathbb{R}^{l^1}$, the learning rate $\eta_1$,  the number of iterations $E_1$.
        \ENSURE  $u_m \in \mathbb{R}^n$, 
        \STATE{Initialize $u_m \in \mathbb{R} ^n$ with all zeros.}
        \FOR{$i = 1, \ldots, E_1$}
        \STATE{$\Delta_{u_m} = (P_m)^T(P_m u_m - d_m) + \rho(u_m - (x_m^{k} - \lambda_m^k / \rho)) $}
        \STATE{$u_m \leftarrow u_m - \eta_1 \Delta_{u_m}$}
        \ENDFOR
	\end{algorithmic}
\end{algorithm} 

Updating $x[m]$ in Eq.~\ref{eq:solve_x} is performed locally as described in Algorithm~\ref{alg:gd_to_solve_x} from current $u_m^{k+1}$, and $\lambda_m^{k}$ where both are stored in memory of the node. We use an iterative approach instead of a direct solution to efficiently handle large-scale data and maintain flexibility in our problem-solving process. By allowing the solution to evolve gradually, the iterative method acts as a regularizer, which can help stabilize the problem and improve performance over time. This approach also enables us to incorporate constraints more effectively and adapt the solution strategy as needed during the iteration process.

After each node computes its respective segment of $x$, these segments must be communicated across all nodes to reconstruct the global $x$. This communication step represents the primary bandwidth demand of the algorithm. However, our approach constrains this demand to a maximum bandwidth that is twice the size of $x$ across the entire network. This includes bandwidth for transmitting each node's segment to all other nodes and for receiving segments from all other nodes.

\begin{algorithm}[t]
	\caption{Gradient Descent to solve Eq.~\ref{eq:solve_x}}
	\label{alg:gd_to_solve_x}
	\begin{algorithmic}[1]
		\renewcommand{\algorithmicrequire}{\textbf{Input:}}
		\renewcommand{\algorithmicensure}{\textbf{Output:}}
		\REQUIRE , $u_m^{k+1}$, $\lambda_m^{k}$, the learning rate $\eta_2$, the number of iteration $E_2$.
        \ENSURE  Locally reconstructed $x[m]^{k+1} \in \mathbb{R}^{n/M}$.
        \FOR{$i = 1, \ldots, E_2$}
        \STATE{$\Delta_{x[m]} = \rho (x[m] - u_m^{k+1} - \lambda_m^k/\rho$)}
        \STATE{$x[m] \leftarrow x[m] - \eta_2 \Delta_{x[m]}$}
        \ENDFOR
	\end{algorithmic}
\end{algorithm}

Finally, updates to the dual variables $\lambda_1, \ldots, \lambda_M$ are carried out using Eq.~\ref{eq:solve_lamda}. This update is performed through in-memory computations, thus requiring no additional communication. We summarize our approach for solving the dADMM problem in Algorithm~\ref{alg:distributed_admm}.

\begin{algorithm}[t]
	\caption{Scalable Distributed tomographic reconstruction (dADMM)}
	\label{alg:distributed_admm}
	\begin{algorithmic}[1]
		\renewcommand{\algorithmicrequire}{\textbf{Input:}}
		\renewcommand{\algorithmicensure}{\textbf{Output:}}
		\REQUIRE Distributed data $d_m \in \mathbb{R}^{l^1}$ and its corresponding matrix $P_m \in \mathbb{R}^{l^1 \times n}$.
        \ENSURE  Global reconstructed $x^{k} \in \mathbb{R}^n$.
         \STATE{Initialize $u^0, x^0, \lambda^0 $ with all zeros}
        \FOR{$k = 1, \ldots, K$}
        \STATE{Update each local $u_m^{k+1}$ by Algorithm~\ref{alg:dis_gd_to_solve_u}}
        \STATE{Update each local $x[m]^{k+1}$ by Algorithm~\ref{alg:gd_to_solve_x}}
        \STATE{Update global $x^{k+1}$ by aggregating all local $x[m]^{k+1}$ }
        \STATE{Update each local $\lambda_m^{k+1}$ by $\lambda_m^{k+1} \leftarrow \lambda_m^k + \rho(u_m - x^{k+1}$)}

        \ENDFOR
	\end{algorithmic}
\end{algorithm}

\subsection{Computing Requirements}

Before introducing the quantization approach for dADMM, we first discuss the memory and communication models relevant to the dADMM algorithm to clarify the concept and the associated communication and processing patterns. Although quantization has not yet been implemented, these models apply to the quantized version with a simple multiplicative constant adjustment based on the compression ratio. Quantization reduces communication bandwidth proportionally to the compression applied, while memory requirements remain unchanged.

\begin{figure}
\centering
  \includegraphics{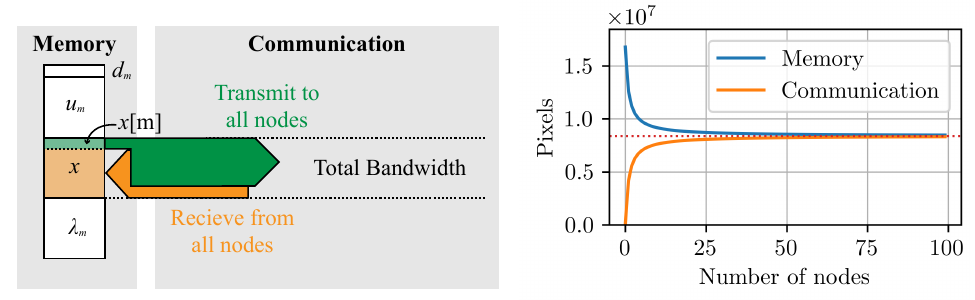}
  \caption{On the left, we illustrate the memory and communication requirements for dADMM within a single node. On the right, we plot these requirements as the number of nodes increases, for a $2048 \times 2048$ tomographic reconstruction image. This includes data with similar dimensions, such as 2048 rotation angles and 2048 detector pixels.}
  \label{fig2}
\end{figure}

Fig.~\ref{fig2} visually illustrates these requirements from the perspective of a representative case for a single node, indexed by $m$. Therefore, the node stores $d_m$, $u_m$, $x$, and $\lambda_m$ in memory and performs in-memory updating them except data at each dADMM iteration. The square blocks represent memory regions for data and unknown parameters. Therefore, the memory requirement for each node is:
\begin{equation}
    \text{Memory}(M) = \frac{D}{M} + 3X,
\end{equation}
where $D$ is the size of the data, $X$ is the size of the image, and $M$ is the number of nodes. As $M$ increases, memory approaches triple the reconstructed image size, $\text{Memory}(\infty) = 3X$, because increasing nodes reduces the data footprint per node. 

The communication overhead, in terms of data sent or received at each node, is modeled as:
\begin{equation}
    \text{Communication}(M) = \frac{2(M - 1)}{M} X.
\end{equation}
The factor $(M - 1)/M$ accounts for each node receiving part of $x$ not available locally and transmitting its part to all other nodes, as illustrated in orange and green in Fig.~\ref{fig2} respectively. Communication also approaches $\text{Communication}(\infty) = 2X$. 

We also illustrate in Fig.~\ref{fig2} the memory and communication requirements as the number of nodes increases, for a representative $2048 \times 2048$ tomographic reconstruction image size. This includes sinogram data with similar dimensions, such as 2048 rotation angles and 2048 detector pixels. These represent minimum requirements, as practical scenarios often require additional memory for solving sub-problems, though estimates can be made based on available solvers.

\subsection{dADMM with Quantization}

In this section, we introduce an extension of dADMM utilizing K-means and JPEG techniques to relax the communication constraints of limited bandwidth across nodes. Our approach closely resembles standard dADMM, distinguished by the integration of quantization within the aggregation problem in Eq.~\ref{eq-compact:conc2}.

In Algorithm~\ref{alg:gd_to_solve_x}, our strategy involves partitioning the vector $x$ across $M$ nodes to distribute its storage and manage the aggregation problem detailed. This partitioning of $x$ enabled scalability. To further optimize communication efficiency, we introduce the quantizer $Q$ on matrices sent to and received from these nodes. This leaves us with the new algorithm as follows:
\begin{align}
\label{eq-compact:solve_u3}
u_m^{k+1} &= \argmin_{u_m} \mathcal{L}_\rho\left(u_m, x^{k}, \lambda_m^{k}\right),
& m=1, 2, \ldots, M, \\ 
\label{eq-compact:solve_x3}
x[m]^{k+1} &= \argmin_{x[m]} \mathcal{L}_\rho\left(u_m^{k+1}, x[m], \lambda_m^{k}\right), & m=1, 2, \ldots, M, \\ 
\label{eq-compact:conc3}
x^{k+1} &= \text{conc}(Q(x[1]^{k+1}), \cdots, Q(x[M]^{k+1})), \\
\label{eq-compact:solve_lambda3}
\lambda_m^{k+1} &= \lambda_m^k + \rho \left(u_m^{k+1}- x^{k+1}\right),
& m=1, 2, \ldots, M,  
\end{align}
where the quantizer $Q(x)$ is applied to $x$ only after solving Eq.~\ref{eq-compact:solve_x3} for aggregation. This strategy aims to reduce communication bandwidth, as illustrated in Fig.~\ref{fig2}, though it come at the cost of precision. To balance effectiveness and efficiency, we employ two distinct quantization techniques: K-means and JPEG. Each technique offers unique advantages, including effectiveness and speed respectively, making them well-suited for our study. Note that K-means will provide a locally-optimal scalar quantizer for the squared-error distortion measure \cite{selim1984k, 7951029}.  

For dADMM with K-means quantization, abbreviated as dADMM-K, each vector designated for transmission across nodes undergoes clustering using the K-means algorithm to determine optimal cluster centers. Subsequently, the original pixel values are updated by assigning each pixel to its closest cluster center. This process is crucial for reducing the amount of data transmitted while preserving essential information. Upon receiving the encoded data, the nodes perform decoding to reconstruct the original vector for further processing and updates. Algorithm~\ref{alg:kmeans on vector} outlines the detailed steps involved in this K-means clustering procedure.

\begin{algorithm}[t]
	\caption{K-means on a vector.}
	\label{alg:kmeans on vector}
	\begin{algorithmic}[1]
 \renewcommand{\algorithmicrequire}{\textbf{Input:}}
	\renewcommand{\algorithmicensure}{\textbf{Output:}}
		\REQUIRE The original vector $\mathbf{a} = [a_1 \cdots a_n] \in \mathbb{R}^{n \times 1}$ is a $n$ dimensional vector.
        \ENSURE  The quantilized $\hat{\mathbf{a}} = [\hat{a_1} \cdots \hat{a_n}] \in \mathbb{R}^{n \times 1}$ is a $n$ dimentinal vector after K-means.

        \STATE{Run K-means on $\mathbf{a} = [a_1 \cdots a_n]$ to get the cluster center $c_1$, $\cdots$, $c_k$, where $k$ is the number of clusters.}

        \FOR{$a_1, a_2, \ldots, a_n$}
        \STATE{$\hat{a_n}$ = $c_k$, which depends on K-means.}
        \ENDFOR
	\end{algorithmic}
\end{algorithm}

Our alternative approach, called dADMM-J, utilizes standard JPEG compression for quantizing vectors during communication. Here, each vector is encoded using the JPEG algorithm before being transmitted across nodes. This method leverages JPEG's efficient compression capacity to minimize data size while maintaining fidelity, thereby enhancing overall communication efficiency.

\section{Results}

We conduct a series of experiments to validate the effectiveness of our dADMM and quantized schemes under various noisy conditions. We begin by assessing the simplest tomographic reconstruction using centralized data, known as Centralized Tomographic Reconstruction (CTR), which serves as a baseline. Next, we address dADMM without quantization, where sinogram features are distributed across multiple nodes, and evaluate the global recovery of the object. Finally, we assess the performance of the qADMM with quantization, using K-means (dADMM-K) and JPEG (dADMM-J) techniques. Performance is assessed using the root mean square error (RMSE) and peak signal-to-noise ratio (PSNR) between the true and reconstructed images, defined as,
\begin{eqnarray}
    \text{RMSE} &=& \left(\frac{1}{n}\sum_{i=1}^n{(X_i - O_i)^2}\right)^{0.5}, \\
    \text{PSNR} &=& 20 \log_{10}\left(\frac{I_{max}}{\mathrm{RMSE}}\right),
\end{eqnarray}
where $X_i$ is the reconstructed image, $O_i$ is the original, $n$ is the total number of pixels, and $I_{max}$ is the maximum image intensity. 

\subsection{Experiment Setup}

\begin{figure}
\centering
\includegraphics{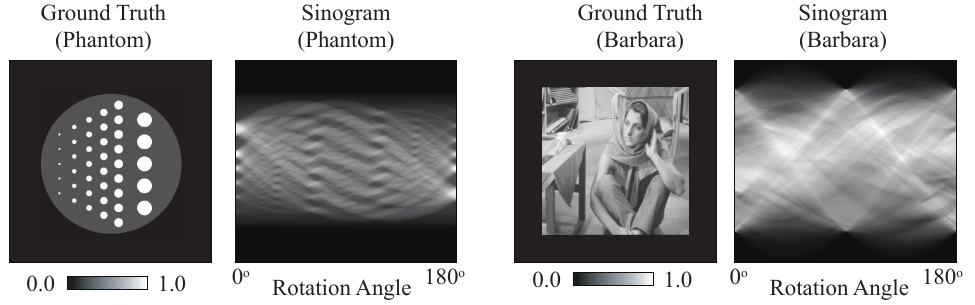}
  \caption{Samples of the true Phantom and Barbara images, along with their corresponding sinograms, used for assessing the performance of the proposed algorithm.}
  \label{fig3}
\end{figure}

For our experiments, we use two datasets: the Phantom image and the Barbara image, each offering unique features, as illustrated in Fig.~\ref{fig3}. The Phantom image, sourced from ToMoBAR \cite{kazantsev2020tomographic}, is a 2D grayscale image with dimensions of $512 \times 512$. To create the sinogram data, we pad the original $512 \times 512$ images to $724 \times 724$ with zeros, leading to a sinogram feature dimension of $724$. Both datasets are acquired using $804$ projection angles, evenly distributed from 0 to 180 degrees, and have 32-bits precision per pixel.

\subsection{Comparison of CTR and dADMM}

\begin{figure}
\centering
  \includegraphics{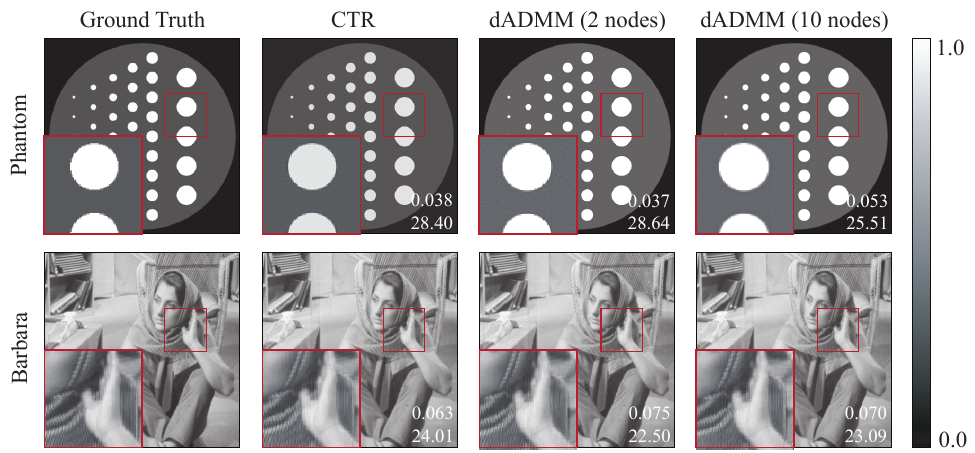}
  \caption{Comparison of reconstructed Phantom and Barbara images using CTR and DTR with 2 and 10 nodes against the ground truth. The values in the bottom left corner of each image indicate the RMSE and PSNR of that image compared to the Ground Truth image shown in the leftmost column.}
  \label{fig4}
\end{figure}

We established CTR as the baseline and compared its performance with dADMM to validate the distributed approach without quantization, as illustrated in Fig.~\ref{fig4}. In dADMM, the sinogram is divided across 2 and 10 nodes, each handling a distinct set of projection angles. For the 2-node configuration, one node processes all odd-indexed angles, while the other manages the even-indexed angles, resulting in each node handling sinogram features corresponding to 402 angles.
In the 10-node configuration, the angles are distributed so that each node ideally covers approximately $804 / 10 = 80.4$ angles. In practice, the first four nodes are assigned 81 angles each, while the remaining six nodes handle 80 angles each, following a sequential index order. For example, node 0 processes indices 0, 10, 20, \ldots, 800, while node 1 handles indices 1, 11, 21, \ldots, 801, and so on.

The local learning rate is set heuristically at $1\times 10^{-6}$, with the aggregation learning rate at $0.2$. For the Phantom data, the number of inner iterations is set to $10$, denoted as $E_1$ in Algorithm \ref{alg:dis_gd_to_solve_u}, and $E_2$ in Algorithm \ref{alg:gd_to_solve_x} is also set to $10$. The number of outer iterations is $1000$. For the Barbara image, due to the complexity of the data, we increase the number of inner iterations, $E1$ and $E2$, to $100$ and set the number of outer iterations to $500$. For both datasets, we use a stopping criterion of $1\times 10^{-6}$ as the convergence threshold, and all variables are initialized to zero before training.

\subsection{Evaluation of dADMM-K and dADMM-J}

In this section, we evaluate the performance of the proposed quantization schemes, dADMM-K and dADMM-J. For dADMM-K, determining the optimal number of clusters ($K$) in the K-means algorithm is required. By clusters, we refer to groups of gray values across the entire image that are sufficiently similar to be represented by a single mean value. A smaller $K$ results in higher compression. For instance, if $K=16$ for an image stored with 32-bit precision, $16=2^4$ bit representation corresponds to a $32/4=8$-fold compression, while no compression occurs when $K=2^{32}$. Although $K$ can be chosen based on the desired compression level, achieving the best balance between compression ratio and image quality requires selecting an optimal value.

\begin{figure}
\centering
\includegraphics{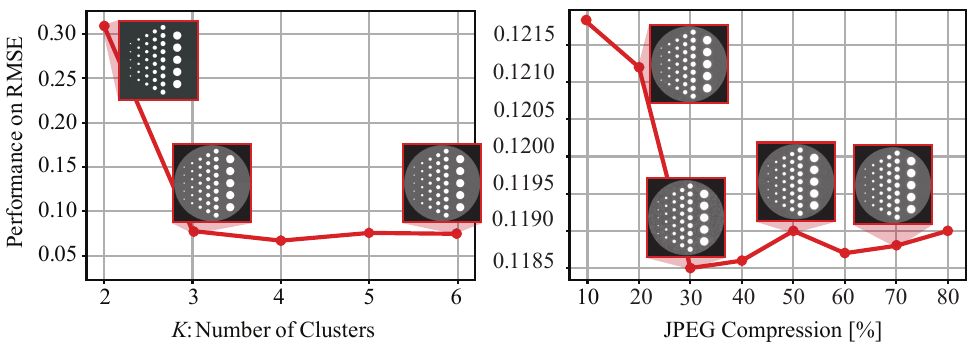}
  \caption{RMSE between the reconstructions and the true image as a function of compression levels for both methods.}
  \label{fig5}
\end{figure}

To this end, we utilized the ``elbow method'' \cite{joshi2013modified}, a conventional approach that identifies the ideal cluster count or $K$ by analyzing the root mean square error (RMSE) for different cluster numbers. The principle behind this method is that if the selected $K$ is smaller than the true number of clusters in the image, the RMSE will be higher. When the number of clusters is too low, the RMSE remains elevated until the chosen value approaches the true number of clusters, causing a noticeable drop. This drop creates an ``elbow'' shape on the RMSE curve, as illustrated in Fig.~\ref{fig5} when $K$ ranges from 2 to 6. The significant reduction in error from $K=2$ to $K=3$ indicates that $K=3$ is optimal for our experiments on the Phantom image. This condition aligns with expectations, as the Phantom image contains three distinct gray levels, making three clusters sufficient for accurate representation, with additional clusters offering diminishing returns. For the Barbara image, the optimal number of clusters was determined to be $K=32$.

We note that these $K$ values of 3 and 32 correspond to using 2-bit and 5-bit precision per pixel, respectively, resulting in compression ratios of $90.6\%$ and $85.3\%$ assuming a 32-bit standard precision for the images.

For JPEG compression, the quality can be set between 0 and 100\%. When plotting the error as a function of compression percentage, it remained centered around 0.12 with minimal variation, and we did not observe a sharp drop indicative of an elbow point. However, a noticeable drop occurred around a 30\% compression ratio, which we used in our calculations.

\begin{figure}
\centering
\includegraphics{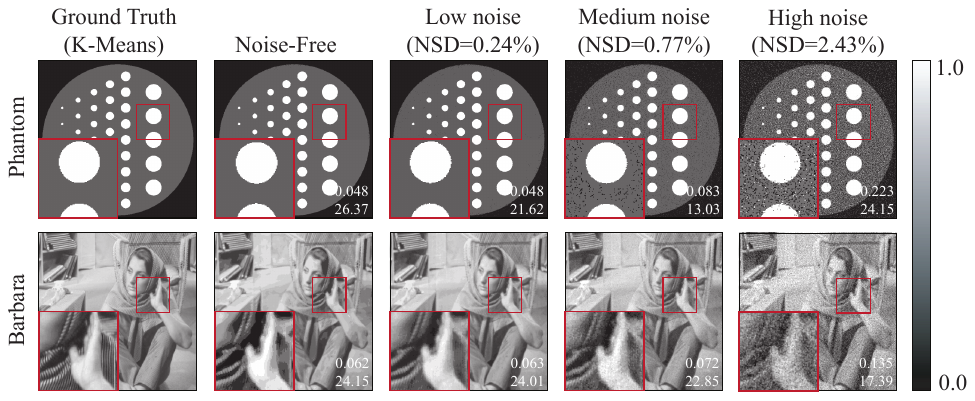}
  \caption{Reconstructions of the Phantom and Barbara images using dADMM-K under varying noise levels, with corresponding RMSE and PSNR values reported.}
  \label{fig6}
\end{figure}

In Fig.~\ref{fig6}, we show the reconstructed images using dADMM-K under noise-free conditions and varying noise levels. Gaussian noise is added to the object's sinogram to simulate different noisy scenarios. To evaluate the algorithm's robustness to noise, we use the Normalized Standard Deviation (NSD), which quantifies the variability of gray values relative to the peak gray value:
\begin{equation}
\text{NSD} = \left(\frac{\sigma}{x_{\text{peak}}}\right) \times 100\%
\end{equation}
where \(x_{\text{peak}} \approx 410\) for both the Phantom and Barbara images. For comparison, we apply K-means with the same \(K\) value as used in the dADMM-K algorithm, with results presented in the leftmost column of Fig.~\ref{fig6}. We evaluate noise performance using NSD values of 0.24\%, 0.77\%, and 2.43\%, representing different levels of noise from low to high.

The Phantom reconstructions were highly accurate under noise-free and low-noise conditions (NSD=0.24\%). However, as the noise level increased (NSD=0.77\%, and 2.43\%), noise began to manifest in the reconstructions. This noise resembled a salt-and-pepper pattern, affecting selected pixels rather than uniformly spreading across all pixels as conventional noise does. This effect likely occurs because a pixel is only assigned to an incorrect cluster when its gray value is significantly altered by noise, making it closer to the mean of a neighboring cluster. Otherwise, the pixel remains in its original cluster, with the overall impact of noise being minimal due to the average noise across the cluster being approximately zero.

Reconstructing the Barbara image, which features more complex details, offers a deeper understanding of the algorithm's performance. The noise effect mirrors that seen with the Phantom image, becoming more noticeable at higher noise levels. However, due to the greater number of clusters in the Barbara image, noise affects nearly all pixels. Interestingly, at a low noise level (NSD=0.24\%), the reconstruction appears closer to the ground truth than the noise-free image. This may be due to a perceptual effect, where the softer isocontours between constant gray levels create a slightly blurred but more natural appearance. In contrast, the noise-free reconstruction has a more cartoonish look, with the isocontours being more pronounced and noticeable. Despite these observations, none of the reconstructions fully capture all the features of the ground truth image. 

\begin{figure}
\centering
\includegraphics{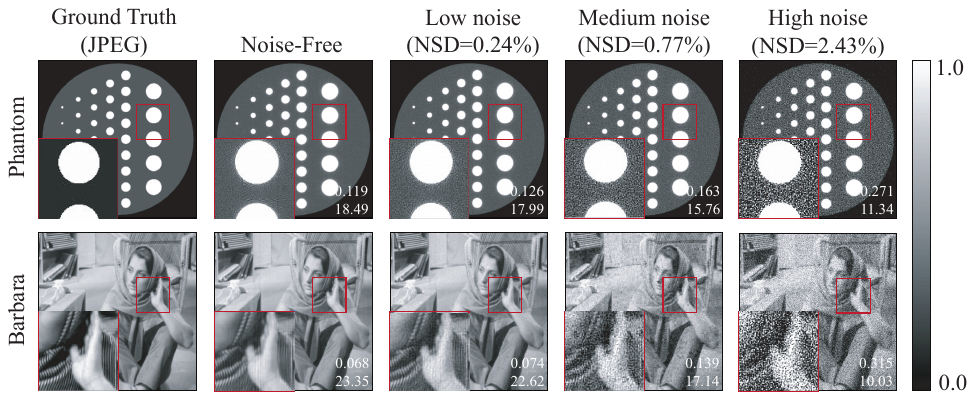}
  \caption{Reconstructions of the Phantom and Barbara images using dADMM-J under varying noise levels, with corresponding RMSE and PSNR values reported.}
  \label{fig7}
\end{figure}

In Fig.~\ref{fig7}, we present the reconstructions of the Phantom and Barbara images using dADMM-J under the same conditions as dADMM-K. Generally, the image quality drops with increasing noise and overall the image quality metrics are poorer than the dADMM-K.

\begin{figure}
\centering
\includegraphics{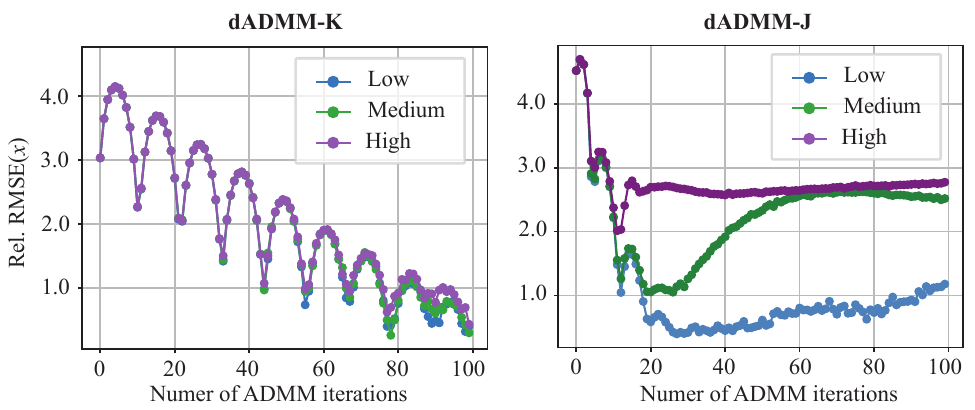}
  \caption{Convergence plots for dADMM-K (left) and dADMM-J (right) during the reconstruction of the Barbara image. The plots show RMSE across successive ADMM iterations between $x$ updates. }
  \label{fig8}
\end{figure}

To study the convergence behavior of the proposed algorithms, we plot the relative RMSE across successive ADMM iterations between $x$ updates in Fig.~\ref{fig8}. On the right, the RMSE plots for dADMM-K exhibit an oscillatory pattern but show a general downward trend across all three noise levels, indicating similar convergence behavior. In contrast, the plots for dADMM-J reach a valley around 15-25 iterations before beginning to increase slightly. This semi-convergence behavior is likely related to the nature of JPEG compression, where the cluster centers struggle to stabilize due to the inherent artifacts introduced by the compression process.  

\section{Discussion}

In this study, we utilized the elbow method to determine the optimal number of clusters for quantization in distributed tomographic reconstruction. This method helps strike a balance between model complexity and performance by identifying an optimal $K$ that balances data compression with reconstruction quality. While effective, the elbow method can be computationally intensive and may not be feasible for every reconstruction due to its cost.

For imaging targets that belong to specific categories (e.g., biological cells, human brains) or well-characterized samples (e.g., microelectronics), pre-determined optimal $K$ values can be established. This approach can simplify the process and provide more accurate and efficient quantization, leveraging prior knowledge to improve reconstruction outcomes. Such targeted methods can streamline the quantization process and enhance overall performance in practical applications.

Incorporating regularization methods into the reconstruction process is another avenue to address noise and improve solution stability. By adding regularization as a separate ADMM sub-problem, we can better manage noise effects, enhance the robustness of the reconstruction, and potentially accelerate convergence. Regularization techniques, such as total variation or Tikhonov regularization, can be particularly effective in mitigating the impact of noise, leading to more stable and accurate reconstructions.

For scenarios involving low photon counts, Bayesian models provide a promising alternative. These models, which account for Poisson counting statistics, offer a better fit for the noise characteristics associated with low photon levels. Bayesian approaches can incorporate prior knowledge about the image or noise distribution, leading to more accurate reconstruction and improved noise handling. This is especially relevant in cases where traditional methods struggle with noise due to the sparse data typical of low photon scenarios.

Overall, our findings emphasize the importance of selecting appropriate methods and models for different reconstruction contexts. The elbow method and regularization techniques play crucial roles in optimizing performance, while Bayesian models offer enhanced capabilities for handling low photon counts. Future work could further explore these strategies and their integration to advance the effectiveness of distributed tomographic reconstruction.

\section{Conclusion} 

We tackle scalable distributed tomographic reconstruction with quantization by employing K-means and JPEG methods, facilitating effective object recovery from distributed data while reducing communication costs. Although our focus is on these two widely used methods, the framework is potentially adaptable to other quantizers. Our results indicate that K-means generally produces smoother and more accurate reconstructions compared to JPEG. Through these exemplar studies, we highlight the trade-offs between reconstruction quality and communication efficiency in quantized distributed tomography.

\section*{Acknowledgements} This research used resources of the Advanced Photon Source, a U.S. Department of Energy (DOE) Office of Science User Facility and is based on work supported by Laboratory Directed Research and Development (LDRD) funding from Argonne National Laboratory, provided by the Director, Office of Science, of the U.S. DOE under Contract No. DE-AC02-06CH11357. R. Miao and E. Koyuncu's work was supported in part by the Army Research Lab (ARL) under Grants W911NF-2120272 and W911NF-2420172, by National Science Foundation (NSF) under Grant CNS-2148182, and by the Army Research Office (ARO) under Grant W911NF-2410049.

\bibliographystyle{ieeetr}
\bibliography{main}

\begin{thebibliography}{10}

\bibitem{kak2001principles}
A.~C. Kak and M.~Slaney, {\em Principles of computerized tomographic imaging}.
\newblock SIAM, 2001.

\bibitem{wernick2004emission}
M.~N. Wernick and J.~N. Aarsvold, {\em Emission tomography: the fundamentals of
  PET and SPECT}.
\newblock Elsevier, 2004.

\bibitem{withers2021x}
P.~J. Withers, C.~Bouman, S.~Carmignato, V.~Cnudde, D.~Grimaldi, C.~K. Hagen,
  E.~Maire, M.~Manley, A.~Du~Plessis, and S.~R. Stock, ``X-ray computed
  tomography,'' {\em Nature Reviews Methods Primers}, vol.~1, no.~1, p.~18,
  2021.

\bibitem{de2018tomobank}
F.~De~Carlo, D.~G{\"u}rsoy, D.~J. Ching, K.~J. Batenburg, W.~Ludwig,
  L.~Mancini, F.~Marone, R.~Mokso, D.~M. Pelt, J.~Sijbers, {\em et~al.},
  ``Tomobank: a tomographic data repository for computational x-ray science,''
  {\em Measurement Science and Technology}, vol.~29, no.~3, p.~034004, 2018.

\bibitem{garcia2021tomoscopy}
F.~Garc{\'\i}a-Moreno, P.~H. Kamm, T.~R. Neu, F.~B{\"u}lk, M.~A. Noack,
  M.~Wegener, N.~von~der Eltz, C.~M. Schlep{\"u}tz, M.~Stampanoni, and
  J.~Banhart, ``Tomoscopy: Time-resolved tomography for dynamic processes in
  materials,'' {\em Advanced Materials}, vol.~33, no.~45, p.~2104659, 2021.

\bibitem{feldkamp1984practical}
L.~A. Feldkamp, L.~C. Davis, and J.~W. Kress, ``Practical cone-beam
  algorithm,'' {\em JOSA A}, vol.~1, no.~6, pp.~612--619, 1984.

\bibitem{katsevich2002theoretically}
A.~Katsevich, ``Theoretically exact filtered backprojection-type inversion
  algorithm for spiral ct,'' {\em SIAM Journal on Applied Mathematics},
  vol.~62, no.~6, pp.~2012--2026, 2002.

\bibitem{andersson2016fast}
F.~Andersson, M.~Carlsson, and V.~V. Nikitin, ``Fast algorithms and efficient
  gpu implementations for the radon transform and the back-projection operator
  represented as convolution operators,'' {\em SIAM Journal on Imaging
  Sciences}, vol.~9, no.~2, pp.~637--664, 2016.

\bibitem{koyuncu2022centroidal}
E.~Koyuncu, ``Centroidal clustering of noisy observations by using th power
  distortion measures,'' {\em IEEE Transactions on Neural Networks and Learning
  Systems}, vol.~35, no.~1, pp.~1430--1438, 2022.

\bibitem{gordon1970art}
R.~Gordon, R.~Bender, and G.~T. Herman, ``Algebraic reconstruction techniques
  (art) for three-dimensional electron microscopy and x-ray photography,'' {\em
  Journal of theoretical biology}, vol.~29, no.~3, pp.~471--481, 1970.

\bibitem{andersen1984simultaneous}
A.~H. Andersen and A.~C. Kak, ``Simultaneous algebraic reconstruction technique
  (sart): a superior implementation of the art algorithm,'' {\em Ultrasonic
  imaging}, vol.~6, no.~1, pp.~81--94, 1984.

\bibitem{gilbert1972iterative}
P.~Gilbert, ``Iterative methods for the three-dimensional reconstruction of an
  object from projections,'' {\em Journal of theoretical biology}, vol.~36,
  no.~1, pp.~105--117, 1972.

\bibitem{hestenes1952methods}
M.~R. Hestenes and E.~Stiefel, ``Methods of conjugate gradients for solving
  linear systems,'' tech. rep., NBS Washington, DC, 1952.

\bibitem{nesterov2013introductory}
Y.~Nesterov, {\em Introductory Lectures on Convex Optimization: A Basic
  Course}.
\newblock Springer Science \& Business Media, 2013.

\bibitem{nocedal2006numerical}
J.~Nocedal and S.~Wright, {\em Numerical Optimization}.
\newblock Springer Science \& Business Media, 2006.

\bibitem{lange1984reconstruction}
K.~Lange and R.~Carson, ``Reconstruction of emission and transmission tomograms
  using maximum likelihood,'' {\em Journal of Computer Assisted Tomography},
  vol.~8, no.~2, pp.~306--316, 1984.

\bibitem{10.1145/3653981}
R.~Miao and E.~Koyuncu, ``Federated momentum contrastive clustering,'' {\em ACM
  Trans. Intell. Syst. Technol.}, vol.~15, June 2024.

\bibitem{10683880}
R.~Miao and E.~Koyuncu, ``Contrastive and non-contrastive strategies for
  federated self-supervised representation learning and deep clustering,'' {\em
  IEEE Journal of Selected Topics in Signal Processing}, pp.~1--16, 2024.

\bibitem{yu2006gpu}
H.~Yu, A.~A. Zamyatin, and G.~Wang, ``Gpu-based iterative tomographic
  reconstruction using the conjugate gradient method,'' {\em International
  Journal of Biomedical Imaging}, 2006.

\bibitem{wang2007performance}
Z.~Wang, H.~Yu, B.~De~Man, A.~Zamyatin, Y.~Krastev, and G.~Wang, ``Performance
  evaluation of iterative tomographic reconstruction algorithms,'' {\em
  Proceedings of SPIE Medical Imaging}, vol.~6510, 2007.

\bibitem{bicer2015rapid}
T.~Bicer, D.~Gursoy, R.~Kettimuthu, F.~De~Carlo, G.~Agrawal, and I.~T. Foster,
  ``Rapid tomographic image reconstruction via large-scale parallelization,''
  in {\em European Conference on Parallel Processing}, pp.~289--302, Springer,
  2015.

\bibitem{cui2013distributed}
J.~Cui, G.~Pratx, B.~Meng, and C.~S. Levin, ``Distributed mlem: An iterative
  tomographic image reconstruction algorithm for distributed memory
  architectures,'' {\em IEEE transactions on medical imaging}, vol.~32, no.~5,
  pp.~957--967, 2013.

\bibitem{chen2021scalable}
P.~Chen, M.~Wahib, X.~Wang, T.~Hirofuchi, H.~Ogawa, A.~Biguri, R.~Boardman,
  T.~Blumensath, and S.~Matsuoka, ``Scalable fbp decomposition for cone-beam ct
  reconstruction,'' in {\em Proceedings of the International Conference for
  High Performance Computing, Networking, Storage and Analysis}, pp.~1--16,
  2021.

\bibitem{xu2020enhanced}
H.~Xu, C.~Li, M.~M. Rahaman, Y.~Yao, Z.~Li, J.~Zhang, F.~Kulwa, X.~Zhao, S.~Qi,
  and Y.~Teng, ``An enhanced framework of generative adversarial networks
  (ef-gans) for environmental microorganism image augmentation with limited
  rotation-invariant training data,'' {\em IEEE Access}, vol.~8,
  pp.~187455--187469, 2020.

\bibitem{hidayetouglu2020petascale}
M.~Hidayeto{\u{g}}lu, T.~Bicer, S.~G. De~Gonzalo, B.~Ren, V.~De~Andrade,
  D.~Gursoy, R.~Kettimuthu, I.~T. Foster, and W.~H. Wen-mei, ``Petascale xct:
  3d image reconstruction with hierarchical communications on multi-gpu
  nodes,'' in {\em SC20: International Conference for High Performance
  Computing, Networking, Storage and Analysis}, pp.~1--13, IEEE, 2020.

\bibitem{hidayetouglu2021memxct}
M.~Hidayeto{\u{g}}lu, T.~Bi{\c{c}}er, S.~G. de~Gonzalo, B.~Ren, D.~G{\"u}rsoy,
  R.~Kettimuthu, I.~T. Foster, and W.-M.~W. Hwu, ``Memxct: design,
  optimization, scaling, and reproducibility of x-ray tomography imaging,''
  {\em IEEE Transactions on Parallel and Distributed Systems}, vol.~33, no.~9,
  pp.~2014--2031, 2021.

\bibitem{9478829}
P.~Li, H.~Seferoglu, V.~R. Dasari, and E.~Koyuncu, ``Model-distributed dnn
  training for memory-constrained edge computing devices,'' in {\em 2021 IEEE
  International Symposium on Local and Metropolitan Area Networks (LANMAN)},
  pp.~1--6, 2021.

\bibitem{10138654}
P.~Li, E.~Koyuncu, and H.~Seferoglu, ``Adaptive and resilient model-distributed
  inference in edge computing systems,'' {\em IEEE Open Journal of the
  Communications Society}, vol.~4, pp.~1263--1273, 2023.

\bibitem{buurlage2019geometric}
J.-W. Buurlage, R.~H. Bisseling, and K.~J. Batenburg, ``A geometric
  partitioning method for distributed tomographic reconstruction,'' {\em
  Parallel Computing}, vol.~81, pp.~104--121, 2019.

\bibitem{biguri2020arbitrarily}
A.~Biguri, R.~Lindroos, R.~Bryll, H.~Towsyfyan, H.~Deyhle, I.~El~khalil
  Harrane, R.~Boardman, M.~Mavrogordato, M.~Dosanjh, S.~Hancock, {\em et~al.},
  ``Arbitrarily large tomography with iterative algorithms on multiple gpus
  using the tigre toolbox,'' {\em Journal of Parallel and Distributed
  Computing}, vol.~146, pp.~52--63, 2020.

\bibitem{boyd2011distributed}
S.~Boyd, N.~Parikh, E.~Chu, B.~Peleato, J.~Eckstein, {\em et~al.},
  ``Distributed optimization and statistical learning via the alternating
  direction method of multipliers,'' {\em Foundations and
  Trends{\textregistered} in Machine learning}, vol.~3, no.~1, pp.~1--122,
  2011.

\bibitem{aslan2019joint}
S.~Aslan, V.~Nikitin, D.~J. Ching, T.~Bicer, S.~Leyffer, and D.~G{\"u}rsoy,
  ``Joint ptycho-tomography reconstruction through alternating direction method
  of multipliers,'' {\em Optics express}, vol.~27, no.~6, pp.~9128--9143, 2019.

\bibitem{nikitin2019photon}
V.~Nikitin, S.~Aslan, Y.~Yao, T.~Bi{\c{c}}er, S.~Leyffer, R.~Mokso, and
  D.~G{\"u}rsoy, ``Photon-limited ptychography of 3d objects via bayesian
  reconstruction,'' {\em OSA Continuum}, vol.~2, no.~10, pp.~2948--2968, 2019.

\bibitem{nikitin2021distributed}
V.~Nikitin, V.~De~Andrade, A.~Slyamov, B.~J. Gould, Y.~Zhang, V.~Sampathkumar,
  N.~Kasthuri, D.~G{\"u}rsoy, and F.~De~Carlo, ``Distributed optimization for
  nonrigid nano-tomography,'' {\em IEEE Transactions on Computational Imaging},
  vol.~7, pp.~272--287, 2021.

\bibitem{majee2022codex}
S.~Majee, S.~Aslan, D.~G{\"u}rsoy, and C.~A. Bouman, ``Codex: a modular
  framework for joint temporal de-blurring and tomographic reconstruction,''
  {\em IEEE Transactions on Computational Imaging}, vol.~8, pp.~666--678, 2022.

\bibitem{aslan2021joint}
S.~Aslan, Z.~Liu, V.~Nikitin, T.~Bicer, S.~Leyffer, and D.~G{\"u}rsoy, ``Joint
  ptycho-tomography with deep generative priors,'' {\em Machine Learning:
  Science and Technology}, vol.~2, no.~4, p.~045017, 2021.

\bibitem{barutcu2021limited}
S.~Barutcu, S.~Aslan, A.~K. Katsaggelos, and D.~G{\"u}rsoy, ``Limited-angle
  computed tomography with deep image and physics priors,'' {\em Scientific
  reports}, vol.~11, no.~1, p.~17740, 2021.

\bibitem{sun2021scalable}
Y.~Sun, Z.~Wu, X.~Xu, B.~Wohlberg, and U.~S. Kamilov, ``Scalable plug-and-play
  admm with convergence guarantees,'' {\em IEEE Transactions on Computational
  Imaging}, vol.~7, pp.~849--863, 2021.

\bibitem{sridhar2020distributed}
V.~Sridhar, X.~Wang, G.~T. Buzzard, and C.~A. Bouman, ``Distributed iterative
  ct reconstruction using multi-agent consensus equilibrium,'' {\em IEEE
  Transactions on Computational Imaging}, vol.~6, pp.~1153--1166, 2020.

\bibitem{tikhonov1977illposed}
A.~N. Tikhonov and V.~Y. Arsenin, {\em Solutions of Ill-Posed Problems}.
\newblock Wiley, New York, 1977.

\bibitem{eckstein1992douglas}
J.~Eckstein and D.~P. Bertsekas, ``On the douglas-rachford splittingmethod and
  the proximal point algorithm for maximal monotone operators,'' {\em Math.
  Programm.}, vol.~55, no.~1--3, pp.~293--318, 1992.

\bibitem{selim1984k}
S.~Z. Selim and M.~A. Ismail, ``K-means-type algorithms: A generalized
  convergence theorem and characterization of local optimality,'' {\em IEEE
  Transactions on pattern analysis and machine intelligence}, no.~1,
  pp.~81--87, 1984.

\bibitem{7951029}
E.~Koyuncu and H.~Jafarkhani, ``On the minimum average distortion of quantizers
  with index-dependent distortion measures,'' {\em IEEE Transactions on Signal
  Processing}, vol.~65, no.~17, pp.~4655--4669, 2017.

\bibitem{kazantsev2020tomographic}
D.~Kazantsev and N.~Wadeson, ``Tomographic model-based reconstruction (tomobar)
  software for high resolution synchrotron x-ray tomography,'' in {\em CT
  Meeting}, vol.~2020, 2020.

\bibitem{joshi2013modified}
K.~D. Joshi and P.~S. Nalwade, ``Modified k-means for better initial cluster
  centres,'' {\em International Journal of Computer Science and Mobile
  Computing}, vol.~2, no.~7, pp.~219--223, 2013.

\end{thebibliography}

\end{document}